\documentclass{ifacconf}
\pdfoutput=1

\usepackage[round]{natbib}
\usepackage{graphicx} 
\usepackage[table,xcdraw]{xcolor}
\usepackage{amsmath}
\usepackage{amsfonts}
\usepackage{algorithmic}
\usepackage[utf8]{inputenc}
\usepackage{soul}
\usepackage[normalem]{ulem}
\usepackage{amssymb}
\usepackage{epstopdf}
\def\url#1{}
\begin{document}
\begin{frontmatter}

\title{Suboptimal multirate MPC for five-level inverters\thanksref{footnoteinfo}}

\thanks[footnoteinfo]{We would like to thank the support by MINECO-Spain and FEDER Funds under projects DPI2016-76493-C3-1-R and DPI2016-75294-C2-1-R.}

\author[First]{Joaquin G. Ordonez} 
\author[First]{Francisco Gordillo}
\author[First]{Pablo Montero-Robina}
\author[First]{Daniel Limon}

\address[First]{Departamento de Ingeniería de Sistemas y Automática, Universidad de Sevilla (e-mail: {jgordonez,gordillo,pmontero1,dlm}@us.es).}

\begin{abstract} 
The application of multilevel converters  to renewable energy systems is a growing topic due to their advantages in energy efficiency. Regarding its control, model predictive control (MPC) has become very appealing  due to its natural consideration of discrete inputs, its optimization capability, and the present-day availability of powerful processing hardware. The main drawback of MPC compared to other control techniques in this field is that the control input is held constant during the sampling period, and it is usually difficult or even impossible to reduce this sampling period because of hardware limitations. For this reason, a multirate MPC algorithm is proposed, which allows to change the control input several times within the sampling period. The optimization problem is simplified and made suboptimal to substantially decrease computational burden. This approach is tested in simulation on a three-phase, five-level diode-clamped converter (DCC) operating in inverted mode with a three-phase resistive load. Results show significant reduction in harmonic distortion at the cost of an increase in the number of commutations with respect to a standard MPC operating at the same sampling period. 
\end{abstract}

\begin{keyword}
current control, five-level diode-clamped converter, model predictive control, three-phase inverters, multirate control, power converters.
\end{keyword}

\end{frontmatter}
%===============================================================================

%================================================== 1
\section{Introduction}

In the last years the significance of power converters is growing due to, among other causes, their application to renewable energy systems. Among this type of electronic devices, multilevel converters present additional advantages as they achieve lower current distortion at the expense of an increase in the circuit configuration and control complexity. From the point of view of control system theory, power converters are switched systems: the variables of the system (currents and voltages) are continuous but the control input  is discrete, since it can only take a finite number of possibilities, namely combinations of the switching devices. Each combination corresponds to an operation mode of the circuit. Most of the power converter control design approaches are based on the consideration of the control inputs as continuous signal in such a way that the controller computes, at each sampling time, real values inside a feasibility interval. A discretization stage is then needed to obtain a discrete sequence of the control signal in such a way that its averaged values are close to the output of the controller. This discretization stage is usually called modulation in this field. The two main families of techniques for modulation are  Carrier-Based Pulse Width Modulation (CB-PWM) and Space Vector Modulation (SVM) \citep{Franquelo2008}. Nevertheless, there exist other techniques that take into account, at the outset, the discrete nature of the control input and, thus, do not use averaged values for it. \cite{Kouro2009} presents model predictive control as one of the most successful approaches in this category, and \cite{vazquez2014model} reviews its applications, but it is not the only one technique in this group \citep{albea2017hybrid}. Several MPC techniques have been presented in recent years, seen in the works of \cite{oikonomou2013model} and \cite{karamanakos2018fixed}.

Model predictive control is very appealing in the  field of power electronics since this approach has several advantages in this type of applications. The consideration of the discrete nature of the control input is one of them since the use of averaged models can be avoided, dodging some issues explored in \cite{perreault1997time}. At the same time, the finite number of possibilities for the control action allows, in some cases, the minimization of the cost function by enumeration of all the feasible cases. Furthermore, the different control objectives can be considered  by adding more terms in the cost function. Nevertheless, this approach also presents an important drawback: the usual MPC algorithm assumes that the control input is held constant during each sampling period while the techniques based on averaged models, which use a modulation phase, such as PWM, allow the change in the control action at any instant during the sampling interval, with the only constraint of fixing in advance the number of commutations that are considered in each sampling period. In this way, an important degree of freedom is lost in the usual MPC approach. This fact has been overcome at the cost of increasing the sampling frequency but the size of the sampling time is a very limiting factor in this field. Nowadays, usual sampling times take values of a few hundreds of microseconds. Reduction of this number complicates the sampling process and may yield computation burden limitations.

The aim of this paper is to present the application of a multirate MPC introduced in \cite{scattolini1994multirate} in power electronics. The feature of multirate MPC is the ability to adjust the control effort several times during one sampling interval. This technique can take the advantages of the MPC approach while improving on the weak point that it has compared with PWM approaches, since multirate MPC allows commutations during the sampling period. 
Simulations compare a standard MPC with the proposed multirate MPC algorithm, and results show the validity of this approach. 

The paper is organized as follows: Section \ref{sec2} presents the five-level diode-clampled inverter, Section \ref{sec3} explains the proposed multirate MPC algorithm, Section \ref{sec4} shows the results of the simulations, and Section \ref{sec5} ends with some conclusions.

%================================================== 2
\section{System description}\label{sec2}

This section presents the converter as well as its modeling required for the control in the following section.

\subsection{Five-level DCC inverter}
\begin{figure}[htb]
	\centering
	\includegraphics[width = 0.95\columnwidth]{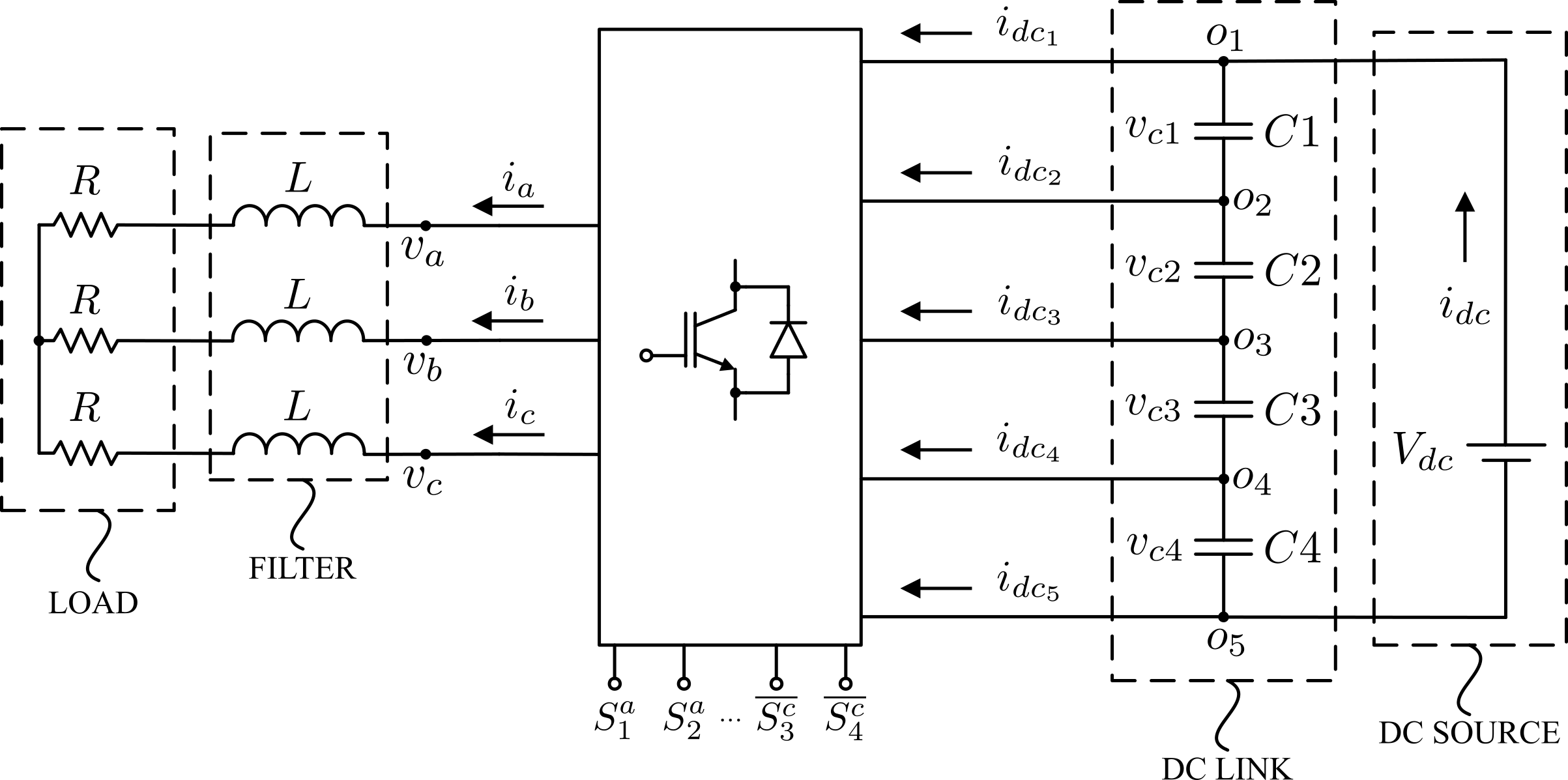}
    \caption{Diagram of the three-phase, five-level DCC inverter.}
    \label{fig:converter}
\end{figure}

\begin{figure}[htb]
	\centering
	\includegraphics[width = 0.95\columnwidth]{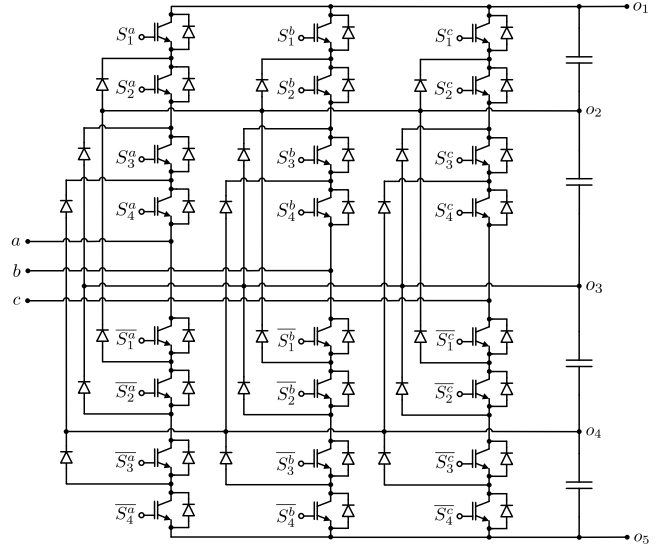}
    \caption{Three-phase, five-level DCC topology.}
    \label{fig:IGBTs}
\end{figure}

The converter considered in this paper is a three-phase, five-level diode-clamped converter  operating in inverter mode connected to a three-phase resistive load  as shown in Fig. \ref{fig:converter}. The central block in this figure represents the set of switching elements depicted in Fig. \ref{fig:IGBTs}. The converter parameters are the inductance of the filter coils $L$, the capacitance of the capacitors that are assumed to be identical $C_1=C_2=C_3=C_4=C$ and the three-phase resistive load value $R$.
%\hl{It is assumed that a resistive load of value $R$ is connected to the dc-link. [En modo inversor, que sentido tiene eso? ademas, esta $R$ puede confundir con la carga trifasica resistiva]}

The control inputs are the state of the switching elements that determine the voltage output of the converter, i.e. the voltage $v_a,v_b,v_c$ measured with respect to the neutral point of the load. Five switching states $f_{ij}$, with $j=1,\dots,5$ are considered for each phase $i$, with $i=a,b,c$ as defined in Tab. \ref{tab:switches}. Variables $S_i^j$ correspond to the switching devices represented in Fig. \ref{fig:IGBTs}. At every instant, for each phase, one and only one $f_{ij}, with j=1,\dots,5$, is equal to 1, while the rest is equal to zero. The last column in Tab. \ref{tab:switches} presents the corresponding voltage output $v_i$ measured between points $i$ and $o_3$ in Fig. \ref{fig:IGBTs}. The previous column shows the five possible values for the control input $u_i$. Taking into account the three phases, it can be seen that the control input can take any of the $5^3=125$ possible combinations.

\begin{table}[htb]
\caption{Feasible combinations for the control input ($i=a,b,c$).}
\centering
\label{tab:switches}
\begin{tabular}{c||c|c|c|c||c|l} 
	$f_{i_j}$ & $S_1^i$ & $S_2^i$ & $S_3^i$ & $S_4^i$ & $u_i$ & $v_i$ \\[0.5mm]
    \hline
    \hline
    $f_{i_1} = 1$	& 0	& 0	& 0	& 0	& -2 & $-v_{c_3}-v_{c_4}$ \\[0.5mm]
    \hline
    $f_{i_2} = 1$	& 0	& 0	& 0	& 1	& -1 & $-v_{c_3}$ \\[0.5mm]
    \hline
    $f_{i_3} = 1$	& 0	& 0	& 1	& 1	& 0 & $0$ \\[0.5mm]
    \hline
    $f_{i_4} = 1$	& 0	& 1	& 1	& 1	& 1 & $v_{c_2}$	\\[0.5mm] 
    \hline
    $f_{i_5} = 1$	& 1	& 1	& 1	& 1	& 2 & $v_{c_1}+v_{c_2}$	\\[0.5mm]
\end{tabular}
\end{table}

\subsection{State-space model}
A model of the five-level DCC will be necessary for the control design to predict the behaviour of the system. %For this paper, the five-level DCC inverter presented in Fig. \ref{fig:converter} is used. 
The dynamic equation for this system is
\begin{equation}\label{contmodel}
    v_i(t) = R \, i_i(t) + L \, \frac{di_i(t)}{dt} \,,
\end{equation}
where $v_i$ is given in Tab. \ref{tab:switches} as a function of the control input $u_i$. Integrating \eqref{contmodel} from $t=kT_s$ to $t=(k+1)T_s$, the model is translated from continuous to discrete time, as explained in \cite{geyer2016}. The system can then be described by a linear time invariant (LTI) state space model in discrete time
\begin{equation}\label{discmodel}
    i(k+1) = A \, i(k) + B \, u(k)
\end{equation}
where $u \in \{-2,-1,0,1,2\}^3$ are the switching states and $i \in \mathbb{R}^3$ are the grid currents. The model parameters are 
\begin{equation}\label{modelparam}
    A=1-\frac{RT_s}{L} \,, \quad B=\frac{V_{dc}T_s}{4L}
\end{equation}
where it has been assumed that the four capacitors are equally charged and, thus, 
$v_{c1}=v_{c2}=v_{c3}=v_{c4}=V_{dc}/4$, being $V_{dc}$ the voltage of the DC source. 

In fact, the voltage balance among capacitors is one of the control objectives, as it is necessary to ensure validity of the model \eqref{discmodel}, \eqref{modelparam}. In order to do that, a model of the capacitors is also required.
The equations that govern these capacitor voltages are
\begin{subequations}\label{cap1}
    \begin{align}
        C \frac{dv_{c_1}}{dt} &= i_{dc} - \sum_{i=\{a,b,c\}} f_{i1} \, i_i \\
        C \frac{dv_{c_2}}{dt} &= i_{dc} - \sum_{i=\{a,b,c\}} (f_{i1}+f_{i2}) \, i_i \\
        C \frac{dv_{c_3}}{dt} &= i_{dc} + \sum_{i=\{a,b,c\}} (f_{i4}+f_{i5}) \, i_i \\
        C \frac{dv_{c_4}}{dt} &= i_{dc} + \sum_{i=\{a,b,c\}} f_{i5} \, i_i \;.
    \end{align}
\end{subequations}
Since $V_{dc} = v_{c_1} + v_{c_2} + v_{c_3} + v_{c_4}$, these four equations can be reduced to three in terms of the difference between the capacitor voltages
\begin{subequations}\label{cap2}
    \begin{align}
        v_{d_1} &= v_{c_1} - v_{c_4} \\ 
        v_{d_2} &= v_{c_2} - v_{c_3} \\ 
        v_{d_3} &= v_{c_3} - v_{c_4}.
    \end{align}
\end{subequations}
In this case, the control objective is to drive these differences to zero. 
Deriving \eqref{cap2} with respect to time and combining with \eqref{cap1} we arrive at the continuous-time model for capacitor balancing
\begin{subequations}
    \begin{align}
        C \frac{dv_{d_1}}{dt} &= \sum_{i=\{a,b,c\}} (-f_{i1}-f_{i5}) \, i_i \label{eq:dvd1}\\
        C \frac{dv_{d_2}}{dt} &= \sum_{i=\{a,b,c\}} (-f_{i1}-f_{i2}-f_{i4}-f_{i5}) \, i_i \label{eq:dvd2}\\
        C \frac{dv_{d_3}}{dt} &= \sum_{i=\{a,b,c\}} f_{i4} \, i_i \;. \label{eq:dvd3}
    \end{align}
\end{subequations}
In discrete time, this model can be written as
\begin{equation}
    v_d(k+1) = v_d(k) +  M(u,k) \cdot i(k+1)
    %dv_d(k) = M(u,k) \cdot i(k)
\end{equation}
where $v_d \in \mathbb{R}^3$, $M : \mathbb{R}^3 \rightarrow \mathbb{R}^{3x3}$, and $i \in \mathbb{R}^3$. The function $M$ is defined as follows
\begin{subequations}
    \begin{align}
        M &= \frac{T_s}{C} [m_a\,m_b\,m_c] \\
        m_i &= \begin{cases}
        \begin{bmatrix}-1 & -1 & 0\end{bmatrix}^\intercal & u_i=-2 \\
        \begin{bmatrix}0 & -1 & 1\end{bmatrix}^\intercal & u_i=-1 \\
        \begin{bmatrix}0 & 0 & 0\end{bmatrix}^\intercal & u_i=0 \\
        \begin{bmatrix}0 & -1 & 0\end{bmatrix}^\intercal & u_i=1 \\
        \begin{bmatrix}-1 & -1 & 0\end{bmatrix}^\intercal & u_i=2 \\
        \end{cases}
    \end{align}
\end{subequations}
with $u_i$ indicating the switching state of phase $i = a, b, c$ as shown in Tab. \ref{tab:switches}.

%================================================== 3
\section{Control design}\label{sec3}

The control scheme follows the concepts of model predictive control (MPC) in \cite{camacho2004}, where a cost function that depends on system variables and several constraints is minimized. Algorithm \ref{alg1} shows the control process of MPC, which is repeated periodically every sampling period $T_s$. 

\begin{algorithm}[htb]
\hrule \vspace{0.1cm}
At each sample time k,
\begin{enumerate}
    \item determine current situation by reading the sensors,
    \item find the optimal sequence of control actions by predicting the behaviour of the system and solving an optimization problem until a control horizon N,
    \item apply the first control action of the sequence corresponding to k.
\end{enumerate}
\vspace{0.1cm} \hrule 
\caption{Model predictive control.}\label{alg1}
\end{algorithm}

When applying MPC in power electronics, achieving lower sampling times leads to a more frequent control action, resulting in increased accuracy at reference tracking. In this case, lower total harmonic distortion (THD). Usually, there exists a practical lower bound for the sampling time due to computation time and measurement frequency limitations. 
In this type of systems, with a relatively low number of possible control inputs (125 in five-level DCCs), it is a common programming practice to check every possible input, calculate the cost function for each of them, and select the one that returns the lower cost. This can be done along a control horizon $N$, which indicates the length of the sequence that is predicted into the future.
The number of tests for this algorithm is $125^N$, which makes the computational burden to grow exponentially with $N$, sometimes making the corresponding computation time too high and non-feasible even for $N=2$. Some techniques have been researched to simplify the computational cost and allow MPC implementations with control horizon $N>1$, as it can be found in \cite{karamanakos2014direct}. Since this matter is not the focus of this paper, $N=1$ will be assumed from now on.

The following optimization problem
\begin{subequations}\label{optprob}
	\begin{align}
	\underset{u}{\min} \quad  & \lambda_I \, |i - i_{ref}| + |u - u_0| \nonumber\\
	& + \lambda_C \, (v_d - v_{d,m}) \, v_{d,m} \label{optprob.o} \\[1.2mm]
	\text{s.t.} \quad 
	&  i = A \, i_0 + B \, u \label{optprob.a}\\
	&  u \in \{-2,-1,0,1,2\}^3 \label{optprob.b}\\
	&  v_d = M(u) \cdot i \label{optprob.c}\\
	&  i_0 = i_m \label{optprob.d}\\
	&  u_0 = u_m \label{optprob.e}
	\end{align}
\end{subequations}
is the core of the MPC algorithm adapted to a five-level DCC. The cost function includes tracking error, number of commutations (which is proportional to commutation losses), and capacitor unbalance, as shown in \eqref{optprob.o}. Weights $\lambda_i$ and $\lambda_c$ adjust the penalization between these three terms. Higher tracking error increases THD, higher number of commutations result in more electric losses, and higher capacitor unbalance reduces the accuracy of the model. The current reference $i_{ref} \in \mathbb{R}^3$ is given by an outer controller in charge of handling the delivered power. The unknown of the optimization problem is the switching state $u \in \mathbb{Z}^3$. The constraints are  the system model \eqref{optprob.a} (which predicts the grid current $i \in \mathbb{R}^3$ for phases $a,b,c$), the control input constraints \eqref{optprob.b}, and the balancing capacitor equations $dv_d \in \mathbb{R}^3$ \eqref{optprob.c}. This problem receives the initial conditions $i_m \in \mathbb{R}^3$ and $v_{d,m} \in \mathbb{R}^3$ from measurements at every sampling instant, and $u_m \in \mathbb{Z}^3$ from the previous optimization problem.

A standard MPC implementation for the optimization problem \eqref{optprob} is described in Algorithm \ref{alg2}. The process is repeated with a frequency of $1/T_s$.

\begin{algorithm}[h]
\hrule \vspace{0.1cm}
At each sample time k,
\begin{enumerate}
    \item update $i_m$ and $v_{d,m}$ from sensor measurement, and $u_m$ from the optimization problem at $k-1$;
    \item solve problem \eqref{optprob} to find the optimal control input;
    \item apply the control action $u$.
\end{enumerate}
\vspace{0.1cm} \hrule 
\caption{Standard MPC.}\label{alg2}
\end{algorithm}

Standard MPC has the drawback of maintaining the control input constant during the sampling period $T_s$. In this work, as an alternative, multirate MPC is presented, derived from standard MPC in order to mitigate this drawback. The multirate MPC technique aims to improve the performance of the converter when the sampling time of the MPC is constrained by the sampling rate of the acquisition system. 

The idea of usual multirate MPC is to allow several commutations during this interval without the need of increasing the sampling rate. When keeping control horizon $N=1$, this method checks $125^{N_\alpha}$ states, with $N_\alpha$ being the number of different values for the control signal allowed during the sampling interval. As before, the computational burden would be too high for any $N_\alpha > 1$, but a simplification is used in this paper to prevent the exponential growth. Instead of optimizing the cost function during the whole sampling interval, the problem can be simplified by optimizing at each subinterval. This might not find the optimal solution for the sampling interval, yielding a suboptimal result, but in this way the computational cost is more affordable. When using the following method, multirate MPC will check $125\times N_\alpha$, and $N_\alpha$ can be increased as long as it is affordable.

\begin{figure}[htb]
	\centering
	\includegraphics[width = 1\columnwidth]{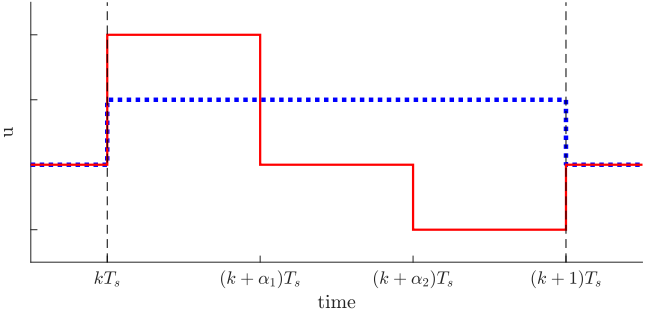}
    \caption{Example of standard MPC (dotted blue plot) and multirate MPC (continuous red plot) control actions with $N_\alpha = 3$ during a sampling period from $kT_s$ to $(k+1)T_s$.}
    \label{fig:mmultirate}
\end{figure}

For any $N_\alpha > 1$, let us define a set of $N_\alpha$ scalars $\alpha_p$, with $p=1,\dots,N_\alpha$, being $0<\alpha_1 <$ $\dots$ $< \alpha_{N_\alpha}$,  and $\alpha_{N_\alpha} = 1$. The sampling interval is split  in $N_\alpha$ subintervals with boundaries $\alpha_1 T_s,\alpha_2 T_s,\dots, \alpha_{N_{ \alpha}-1}T_s$. An example for $N_\alpha = 3$ is illustrated in Fig. \ref{fig:mmultirate}. The selection of $N_\alpha$ depends on computation time restrictions. %Note that the components of $\alpha_p$ vector do not have to split the sampling interval into equal pieces. Their positions do not affect computation time by themselves since the algorithm can be completely computed using only the sensor data at time $k$, and does not wait for new measurements during the interval. Only the number of subintervals $N_\alpha$ will significantly determine the computation time.

Since the previous discrete-time state space model \eqref{discmodel} was obtained for a time period $T_s$, and multirate MPC splits the period in smaller subintervals, the matrices $A$ and $B$ change. The new model for multirate MPC that depends on $\alpha_p$ is
\begin{equation}
    i(p+1)= A_p \, i(p) + B_p \, u(p) \;,
\end{equation}
%\begin{equation}
%    A_p=1-\frac{R\,\alpha_pT_s}{L} \;, \quad B_p=\frac{V_{dc}\,\alpha_pT_s}{4L}\;.
%\end{equation}
\begin{equation}
    A_p=1-\frac{R\,(\alpha_p-\alpha_{p-1})T_s}{L} \;, \quad B_p=\frac{V_{dc}\,(\alpha_p- \alpha_{p-1})T_s}{4L}\;,
\end{equation}
being $p=1,\dots,N_\alpha$ and $\alpha_0=0$.

This approach slightly changes the optimization problem for the multirate MPC into
\begin{subequations}\label{optprob1}
	\begin{align}
	\underset{u(p)}{\min} \quad  & \lambda_I \, |i(p+1) - i_{ref}| + |u(p) - u(p-1)| \nonumber \\
	& + \lambda_C \, (v_d(p+1) - v_{d,m}) \, v_{d,m} \label{optprob1.o} \\[1.2mm]
	\text{s.t.} \quad 
	&  i(p+1) = A_p \, i(p) + B_p \, u(p) \label{optprob1.a}\\
	&  u(p) \in \{-2,-1,0,1,2\}^3 \label{optprob1.b} \\
	&  v_d(p+1) = M(u(p)) \cdot i(p+1) \label{optprob1.c}\;.
	\end{align}
\end{subequations}

Algorithm \ref{alg3} proposes the multirate MPC that solves the optimization problem \eqref{optprob1} $N_\alpha$ times.

\begin{algorithm}[h]
\hrule \vspace{0.1cm}
At each sample time k,
\begin{enumerate}
    \item update $i(1) = i_m$ and $v_{d,m}$ from sensor measurement, and $u(0) = u_m$ from the optimization problem at $k-1$;
    \item $
    \mathbf{for} \; p = 1,\dots,N_\alpha \\ 
    \indent \text{solve problem} \; \eqref{optprob1} \\
    \mathbf{end} ;
    $
    \item apply $u(1),\dots,u(N_\alpha)$ at times $(k+\alpha_0T_s),\dots,(k+\alpha_{N_\alpha-1}T_s)$ respectively.
\end{enumerate}
\vspace{0.1cm} \hrule 
\caption{Proposed multirate MPC.}\label{alg3}
\end{algorithm}

This problem provides a suboptimal solution. As it has been said before, optimality is sacrified in order to obtain a solution in affordable computation time. Also, notice that the optimization problem that can be formulated at time $kT_s$ with the information available at that instant. For that reason, the components of $\alpha_p$ vector do not have to split the sampling interval into equal pieces. Their positions do not affect computation time by themselves since the algorithm can be completely computed using only the sensor data at time $kT_s$, and does not wait for new measurements during the interval. Only the number of subintervals $N_\alpha$ will significantly determine the computation time.

It is worth noting that in all sampling periods, the initial parameters $i(p)$ and $u(p-1)$ for the optimization problem \eqref{optprob1} in Algorithm \ref{alg3} will come once from sensor measurement (always $i(1)$ and $u(0)$), and the remaining $N_\alpha-1$ times (up to $i(N_\alpha)$ and $u(N_\alpha-1)$) from the prediction model.

%================================================== 4
\section{Simulation results}\label{sec4}

In this section, the proposed multirate MPC will be compared in simulations to the standard MPC with control horizon $N=1$. The schematic used for such simulations is presented in Fig. \ref{fig:converter}. The system parameters are the ones shown in Table \ref{tab:systemparam}. The simulations are performed such that a grid current reference of $12$ A of amplitude per phase at $50$ Hz is tracked. 

\begin{table}[hbt]
    \caption{System parameters.}
    \label{tab:systemparam}
    \centering
    \begin{tabular}{c|c|c}
        Variable & Description & Value \\[0.5mm]
        \hline
        \hline
        $R$ & Grid resistive load & $30$ $\Omega$ \\[0.5mm]
        \hline
        $L$ & Filter inductance & $5$ mH \\[0.5mm]
        \hline
        $V_{dc}$ & DC-link voltage & $750$ V \\[0.5mm]
        \hline
        $T_s$ & Sampling period & $20$ $\mu$s \\[0.5mm]
        \hline
        $N_\alpha$ & Number of subintervals & $3$ \\[0.5mm]
        \hline
        $\alpha_p$ & Subinterval fractions & $0.45,0.75,1$ \\[0.5mm]
        \hline
        $\lambda_I$ & Tracking cost & $1\cdot 10^2$ \\[0.5mm]
        \hline
        $\lambda_C$ & Balancing cost & $2\cdot 10^{-4}$ \\[0.5mm]
    \end{tabular}
\end{table}

\begin{figure}[htb]
	\centering
	\includegraphics[width = 1\columnwidth]{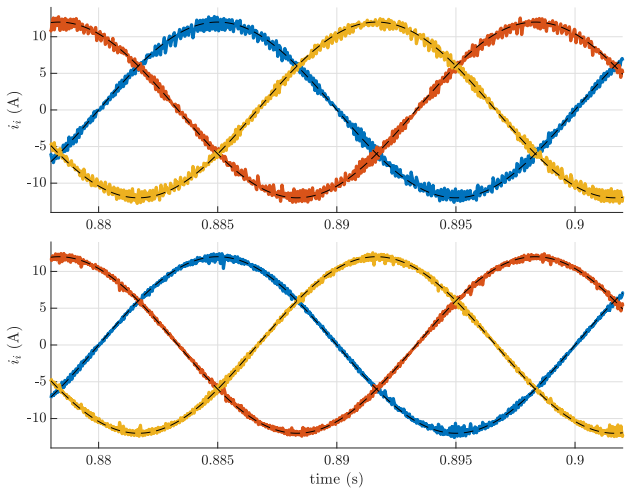}
    \caption{Grid currents tracking a reference of 12 $A$ of amplitude. (Top) Standard MPC. (Bottom) Multirate MPC.}
    \label{fig:mpcsim1}
\end{figure}

\begin{figure}[ht]
	\centering
	\includegraphics[width = 1\columnwidth]{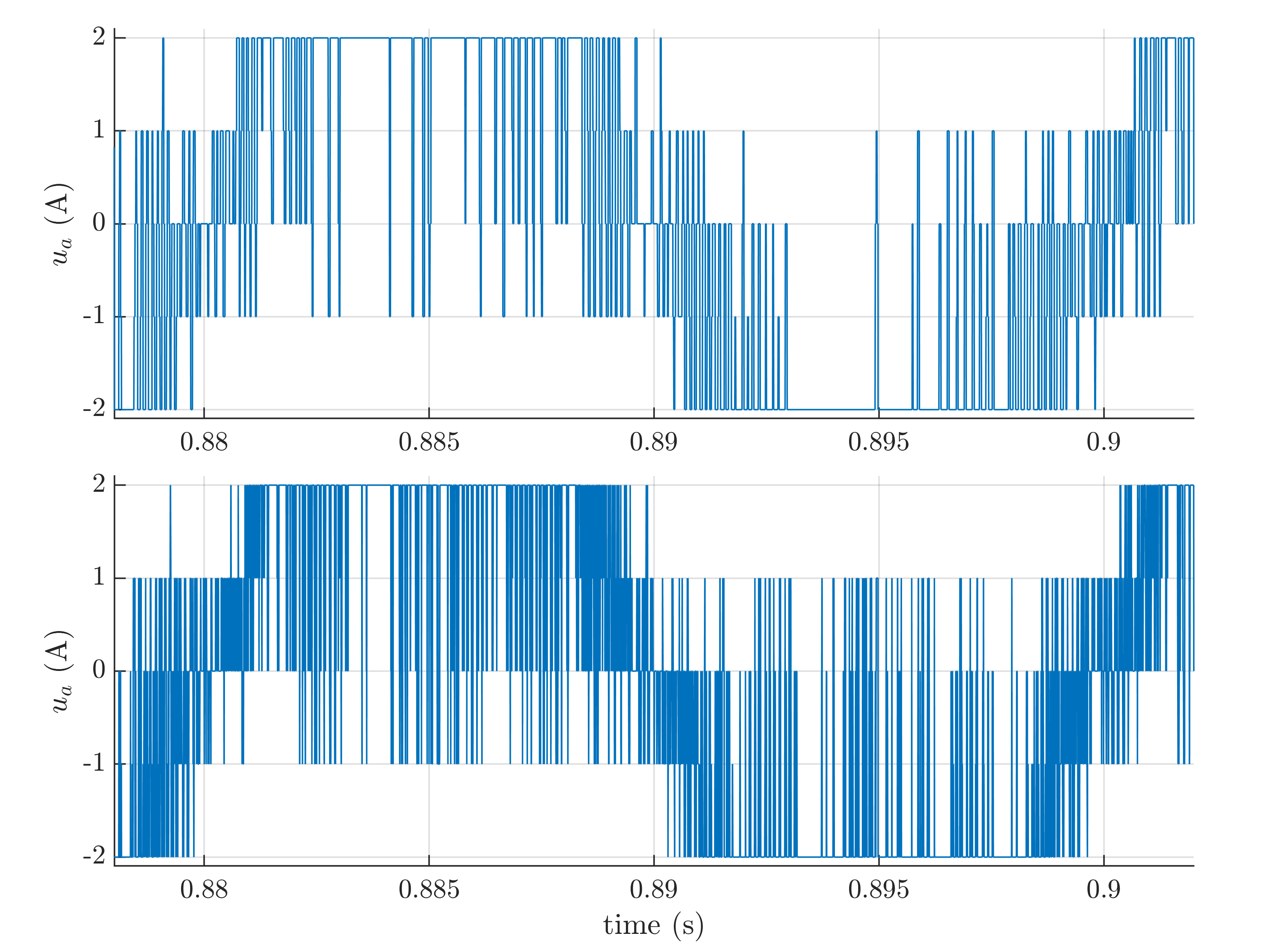}
    \caption{Switching states of the output at steady state. (Top) Standard MPC. (Bottom) Multirate MPC.}
    \label{fig:mpcsim2}
\end{figure}

Figure \ref{fig:mpcsim1} depicts the grid currents achieved at steady-state for the standard MPC and the multirate MPC. It can be seen that the proposed algorithm considerably improves the ripple of the currents. However, this improvement comes at the cost of increasing the amount of commutations. In this regard, Fig. \ref{fig:mpcsim2} shows the normalized output voltage for each of the modulation approaches where this fact is made clear. The average number of commutations obtained in a grid period is 2083 for the multirate MPC and 456 for the standard MPC. Nevertheless, control complexity and implementation have been barely increased, the sampling rate is kept the same and the computation is still performed at the same rate than the standard MPC. Thus, this strategy could be easily implemented to improve the grid current distortion without the need to go for more powerful digital devices or the need to increase the sampling ratio, which could require some hardware upgrades. 

\begin{figure}[ht]
	\centering
	\includegraphics[width = 1\columnwidth]{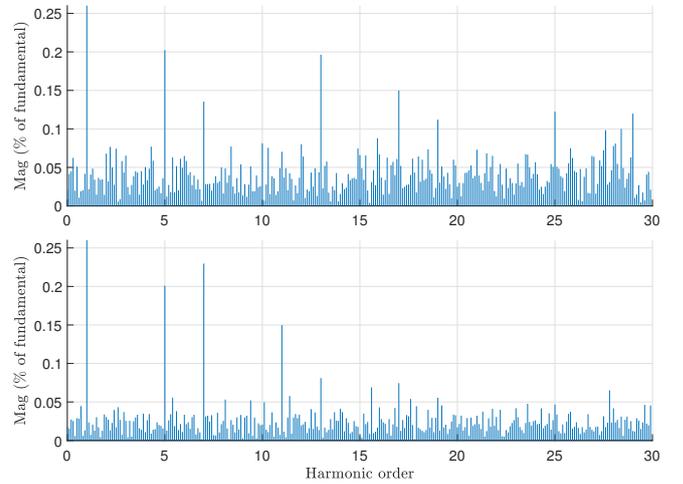}
    \caption{Harmonic spectrum of the grid current related to the fundamental component. (Top) Standard MPC. (Bottom) Multirate MPC.}
    \label{fig:mpcsim4}
\end{figure}

Notice that, because of the nature of the multirate MPC, the harmonic spectrum of the currents are transferred from the low-order harmonics to the high-order ones. This is due to the fact that multirate MPC makes the cost function to yield lower values when additional switching actions are included, and this is done at the sampling frequency rate. Thus, the sampling frequency harmonic and its multiples are increased at the same time lower harmonics are reduced. This is very profitable for the overall performance as the filter rejects more severely the high-frequency harmonics. Figure \ref{fig:mpcsim4} shows the low-order harmonic spectrum of the grid currents at steady state for both approaches. It can be seen that the overall amplitude of low-order harmonics is reduced when multirate MPC is used. At the same time, the harmonics corresponding to frequencies $\{1/T_s,1/2T_s,1/3T_s,\dots\}$ are increased due to the incorporation of additional commutations. As a result, the total harmonic distorsion (THD) measured in Fig. \ref{fig:mpcsim1} is $4.53\%$ for the standard MPC implementation, and $2.52\%$ for the multirate MPC.

\begin{figure}[htb]
	\centering
	\includegraphics[width = 1\columnwidth]{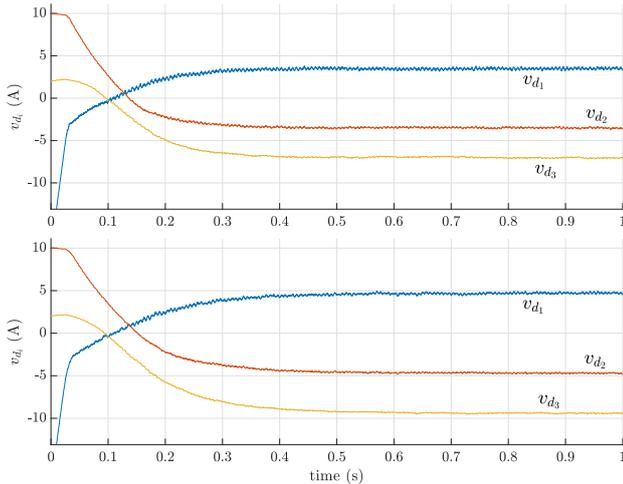}
    \caption{Capacitor voltage differences starting from an unbalanced condition. (Top) Standard MPC. (Bottom) Multirate MPC.}
    \label{fig:mpcsim3}
\end{figure}

Regarding balancing capabilities, Fig. \ref{fig:mpcsim3} depicts the capacitor voltage differences $\{v_{d_1}, v_{d_2}, v_{d_3} \}$ starting from an unbalanced situation for both approaches. It can be seen that they exhibit almost identical behaviour and the three differences are kept close to zero. Consider that these differences can be neglected when compared to the capacitor voltage values and, therefore, the assumption of equal capacitor voltages can remain valid. These differences can be further reduced if the balancing weight $\lambda_C$ in the cost function is increased. This, however, would come at the cost of sacrificing current tracking, and consequently increasing harmonic distortion.

%Para Pablo: las gráficas de arriba son con implementación mpc clásica, las de abajo con el multirate mpc; los thd son de 4.53\% y 2.52\%, medidos desde 0.7 a 0.9 segundos (10 ciclos); el número de conmutaciones son 456 y 2083, medidos desde 0.8 a 0.9 segundos (5 ciclos).

%================================================== 5
\section{Conclusions}\label{sec5}
In this paper, a multirate MPC algorithm has been proposed for a three-phase, five-level inverter. The optimization problem to be solved in each sampling instant has been simplified with the aim of keeping practicable the  computational burden. 
The proposed algorithm has been compared, through simulations, with a standard MPC operating at the same sampling period.
The results show a significant reduction in harmonic distortion at the cost of an increase in the number of commutations. The tradeoff between harmonic reduction and commutations increment can be tuned by means of the choice of the number of subintervals considered by the algorithm. %The proposed technique is of special interest for those systems whose sampling rate limits the potential improvement of the distortion performance.
%\sout{The proposed technique is most recommended when the performance of the system is limited by the sampling rate of the hardware.}

%\pagebreak
%==================================================

\bibliography{ifacbib}

\end{document}